\documentstyle[multicol,aps,prl,epsf,psfig]{revtex}
\def\be{\begin{equation}}
\def\ee{\end{equation}}
\def\bea{\begin{eqnarray}}
\def\eea{\end{eqnarray}}
\def\ba{\begin{array}}
\def\ea{\end{array}}
\def\bc{\begin{center}}
\def\ec{\end{center}}

\def\ri{{\bf r}_i}
\def\Ri{{\bf R}_i}

\def\R{{\bf R}}
\def\DE{\Delta E}

\begin{document}

\title{Glassy Dynamics of Protein Folding} 
\author{Erkan T\"uzel$^1$, Ay{\c s}e Erzan$^{1,2}$}
\address{$^1$  Department of Physics, Faculty of  Sciences
and Letters\\
Istanbul Technical University, Maslak 80626, Istanbul, Turkey\\
$^2$  Feza G\"ursey Institute, P. O. Box 6, \c
Cengelk\"oy 81220, Istanbul, Turkey}
\date{\today}
\maketitle
\begin{abstract}
A coarse grained model of a random polypeptide chain, with only discrete torsional 
degrees of freedom and Hookean springs connecting pairs of hydrophobic residues
is shown to display stretched exponential relaxation under Metropolis dynamics 
at low temperatures with the exponent $\beta\simeq 1/4$,  in agreement with the best 
experimental results. The time dependent correlation functions  for 
fluctuations about the native state, computed in the Gaussian approximation  for  
 real proteins,  have also been found to  have the same functional form. 
Our results indicate that the energy landscape  exhibits  universal features over a very 
large range of energies and  is relatively independent of the specific dynamics.

 PACS No. 87.17.Aa, 5.70.Ln, 64.70.Pf, 82.20.Rp 

\end{abstract}
\begin{multicols}{2}
A huge amount of effort has recently been invested in modeling the
interactions responsible for yielding the native states of proteins
as their thermodynamic equilibrium state~\cite{Wolynes1,Wolynes2}.
It has recently begun to be appreciated
that such features of real proteins as the density of vibrational energy
states~\cite{ben-Avraham} may be reproduced by coarse-grained model hamiltonians
which capture the essential mechanism driving the folding process, namely 
hydrophobic 
interactions~\cite{ben-Avraham,Dill,Tirion,Bahar,Erman1,Erman2,Tuzel}.
In this paper we introduce and study a model of N coupled,
over--damped torsional degrees of freedom with discrete allowed states.
Under Metropolis Monte Carlo dynamics, with random initial conditions, 
we find that at low temperatures the model exhibits 
power law relaxation for the initial stages of decay, and at the later stages
the relaxation obeys a stretched exponential with the exponent $\beta\simeq 
1/4$. 
This type of relaxation behaviour is of the Kohlrausch-Williams-Watts type as
observed experimentally for real proteins 
~\cite{Wolynes1,Angell,Bahar1,Colmenero}.
We find that at zero temperature the probability distribution function of the
energy steps encountered along a relaxation path in phase space also obeys a 
stretched
exponential form, with another exponent $\alpha\simeq 0.39$. We show that
$\beta=\alpha / (\alpha + 1)$, yielding a value for $\beta$ which is in very 
good
agreement with our simulation results. 

We take as our point of departure the model proposed by 
Halilo{\u g}lu, Bahar, Erman~\cite{Bahar}. 
The central idea of this model is that all interactions in the protein
are governed by confining square-law potentials, so all attractions
may be treated as if the residues interact with each
other through Hookean forces ~\cite{Bahar,Erman1,Erman2}.

To keep our model very simple, we 
consider covalent bonds as fixed rods of equal length. The residues located 
at the vertices may be polar $P$ or hydrophobic $H$. All the hydrophobic 
vertices are to be
connected to each other with springs of equal 
stiffness.
This feature mimicks the effective pressure that is exerted on the hydrophobic
residues by the ambient water molecules, and results in their being driven to 
the
relatively less exposed center of the molecule in the low lying energy
states, whereas the polar residues are closer to the surface (see Fig. 1), 
a feature that is common to the native configurations. The constraints
placed on the conformations due to the rigid chemical bond lengths and
restriction of the chemical and dihedral angles to discrete values prevent
the molecule from collapsing to a point.
\begin{figure}
\begin{center}
\leavevmode
\psfig{figure=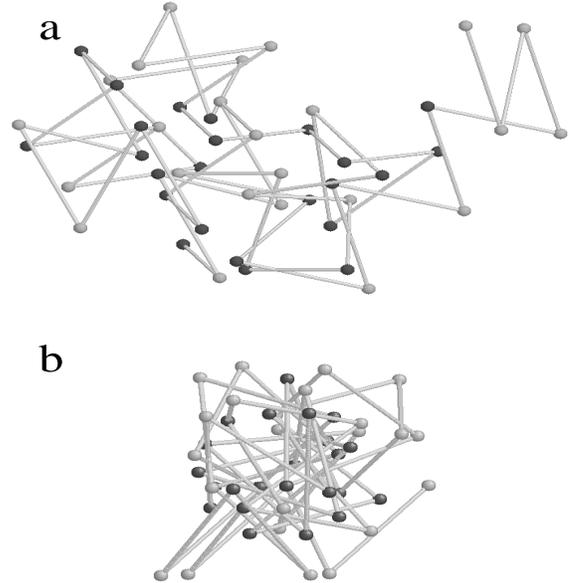,width=8cm,height=9cm,angle=0}
\end{center}
\narrowtext 
\caption{
 A chain of $N=48$ residues, half of which are randomly chosen to be
hydrophobic, (darker beads) 
shown a) in a random initial configuration and b) in a folded state
reached under Metropolis 
dynamics. The chain has folded in such a way as to leave the polar 
residues on the outside. (Generated using RasMol V2.6)}
\label{fig1}
\end{figure}
It is known that real proteins are distinguished by H-~P sequences that
lead to 
unique ground states while a randomly chosen H-P sequence will typically give 
rise to a highly degenerate ground state. Nevertheless, in our Monte Carlo study we 
considered a generic H-P sequence obtained by choosing fifty percent of the residues
to be 
hydrophobic and distributing them randomly along the chain. In the absence of 
detailed knowledge regarding the rules singling out the realistic H-P sequences 
we believed this to be in keeping with our statistical approach.
It might be speculated that
the choice of equal probabilities 
for encountering H and P groups along the chain, and distributing them randomly, 
maximizes the configurational 
entropy of the chain~\cite{Tang} and enhances the ``designability'' giving rise 
to rather realistic results.
\begin{figure}
\begin{center}
\leavevmode   
\psfig{figure=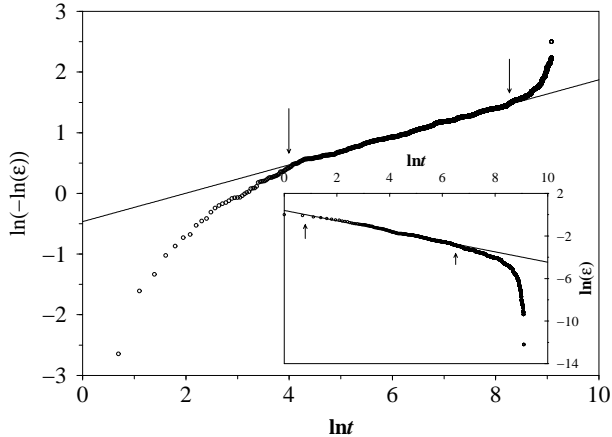,width=8cm,height=6cm,angle=270}
\end{center}
\narrowtext 
\caption{  
 The decay of the energy of an $N=100$ chain from a random initial
configuration,
along a zero temperature Metropolis trajectory, of ~10,000 steps,
averaged over 20 runs.
The later stages fit on a
stretched exponential curve $\epsilon(t) \sim \exp (-t^\beta)$
with $\beta = 0.234 \pm 0.003$.
The initial stage (inset)
is fit by a power law $\epsilon(t) \sim t^{-\sigma}$ with $\sigma = 0.49 \pm
0.01$.}
\label{fig2}   
\end{figure}
Our model for the protein chain consists of $N$ ``residues''
which are treated as point vertices, connected to each other by
rigid rods. The ``bond angle'' $\alpha_i$ at the $i$'th vertex,
$i=1, \ldots, N-1$, is
fixed to be $(-1)^i \alpha,$ with $\alpha=68^\circ$. The
dihedral angles $\phi_i$ can take on the values of 0 and $\pm 2\pi/3 $.
The state (conformation) of the system
is uniquely specified once the numbers $\{\phi_i\}$ are given.
Thus, the residues effectively reside on the vertices of a tetrahedral
lattice.

The energy of the molecule is
\be
E= {K\over 2} \sum_{i,j} c_{i,j} \vert {\bf r}_i-{\bf r}_j\vert^2 
= K \sum_{i,j} {\bf r}_i^{\dagger} V_{ij} {\bf r}_j \label{energy}
\ee
If we define $Q_i=1$ for the $i$' th vertex being occupied by a hydrophobic 
residue, and $Q_i=0$ otherwise, we may write $c_{i,j}=Q_i Q_j$ and  
\begin{eqnarray}
V_{ij}=[(&N&_H-1) c_{i,i} -c_{i,j-1}-c_{i,j+1}]\delta_{i,j} \cr
& -& (1-\delta_{i,j})(1-\delta_{i,j-1}-\delta_{i,j+1}) c_{i,j} \;\;\;.
\end{eqnarray}
\noindent
The position vectors ${\bf r}_i$ of each of the vertices in the
chain can be expressed in terms of a sum over the directors $\Ri$ of unit length 
representing the chemical bonds, which may be obtained from ${\bf R}_1$ by 
successive rotations ${\bf M}_k(\alpha_k)$ 
and ${\bf T}_k(\phi_k)$ through the bond and the dihedral
angles~\cite{Flory},
\be
\ri= \sum_{j=1}^{i-1} \prod_{k=j}^{2} {\bf T}_k(\phi_k) {\bf M}_k(\alpha_k) 
\R_1\;\;\;,
\ee
where we may choose $\R_1$ along any Cartesian direction in our laboratory
frame without loss of generality. 
\begin{figure}
\begin{center}
\leavevmode   
\psfig{figure=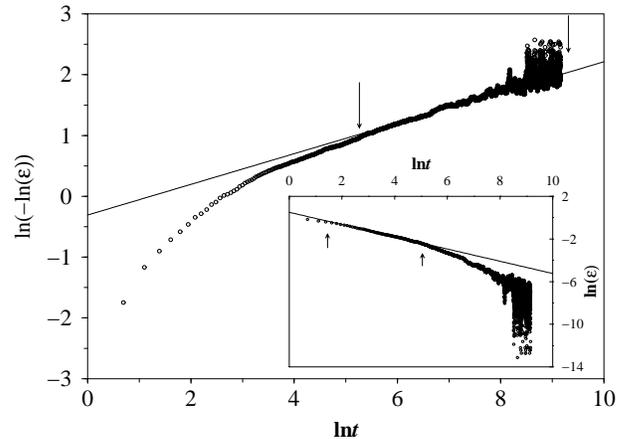,width=8cm,height=6cm,angle=270}
\end{center}  
\narrowtext
\caption{ The decay of the energy of an $N=48$ chain, along a Metropolis
trajectory,
from a random initial configuration averaged over 100 runs for 
$\gamma = 0.3$. 
The initial stage (inset) is fit by a power law
$\epsilon(t) \sim t^{-\sigma}$ with $\sigma = 0.57 \pm 0.01 $, 
and the late stage to a stretched exponential with $\beta = 0.25 \pm 0.03$.}
\label{fig3}
\end{figure}
In order to investigate the relaxation properties of the present
model, we have employed Metropolis Monte Carlo dynamics.  
This consisted of {\it a)} choosing a pair $(i,i^\prime)$ of dihedral angles 
randomly on the chain, and updating the ($\phi_i$, $\phi_{i^\prime}$) in a way 
that preserves angular momentum, incrementing them in opposite directions by 
$\Delta \phi = \pm 2 \pi /3$, {\it b)} accepting the move with unit probability if 
$\Delta E\le 0$ and with probability $p=\exp (-\gamma \Delta E))$ for $\Delta 
E >0$, {\it c)} repeating the second step once before discarding the pair 
altogether and going to the first step. Here $\gamma$ serves as an effective 
inverse temperature. We monitor the relaxation of the total energy 
as a function of ``time'' measured in the number of MC steps, ( i.e.,
the number of pairs $(i,i^\prime)$ sampled) until  a steady state is reached,
typically in about 10,000 steps.
The results for chains of $N=100$ averaged over 20 
randomly chosen initial configurations  at zero temperature ($\gamma=\infty$)
are shown in Fig. 2. Defining
$\epsilon \equiv(E-E_0)/E_I$, where $E_0$ is the (time- averaged) equilibrium 
energy and $E_I$, the initial value, we find that it 
obeys a power law, $\epsilon(t)\sim t^{-\sigma}$
with $\sigma=0.49 \pm 0.01$ for
the 
initial stages of the decay, while later stages can be fitted by a stretched 
exponential $\epsilon(t) \sim e^{-t^{\beta}}$ with  $\beta = 0.234 \pm 0.003$.
We also performed simulations for different values of $\gamma$,
for chains of $N=48$,
averaging over 100 runs with random initial configurations.
For $\gamma \rightarrow
\infty$, $\gamma = 0.5$ 
and $\gamma = 0.3$, the above relaxation behaviour continues to
hold and the exponents
do not seem to depend 
on $\gamma$, with  $\beta \simeq 1/4$ and $\sigma \simeq 1/2$
as given in Table I. 

{\small
{\bf Table I} ~~The exponent $\sigma$ and $\beta$ found for the power law
and stretched exponential decay of the total energy with time, 
for different chain lengths $N$ and 
inverse temperatures $\gamma$. The fits were obtained from a weighted 
least-squares computation.}
\bc
\begin{tabular}{|c|c|c|c|c|c|}
\hline 
$N$ & { $\gamma$} & { $\sigma$} & $\Delta \sigma$ & { $\beta$} & $\Delta
\beta$ \\ \hline
48   &$\infty$ &     0.57 &   0.01   &0.281   &     0.004  \\
   &  0.5      &     0.56 &   0.01      &  0.30  &    0.04  \\
   &  0.3      &     0.57 &   0.01      &  0.25  &    0.03  \\     
\hline
100 & $\infty$ &     0.49 &   0.01     &  0.234    & 0.003   \\ 
\hline
\end{tabular}  
\ec
The variation of the total energy in time is sketched in Fig. 4 over a short 
sequence of relaxation events. Clearly one may write $E(t)$, averaged over many 
independent runs, as $\langle E(t)\rangle = \langle E(0)-\sum_{i=1}^M \Delta E_i 
\Theta (t-t_i) \rangle$
where $\Theta$ is the Heavyside step function and $t_i=\sum_{k=0}^{i-1} \tau_k$. 
Taking the time derivative one gets,
\be
\langle \dot{E}(t) \rangle = \langle -\sum_{i=1}^M \Delta E_i \delta(t-
\sum_{k=0}^{i-1} \tau_k) \rangle \;\;\;. 
\ee
\noindent
At zero temperature, the expectation value of $\dot{E}(t)$ can be calculated by
carrying out an 
integration over the distibution of waiting times $\{ \tau_k \}$, and the 
distribution of energy steps encountered along the relaxation path. 
The expectation value, $\langle \dot{E}(t) \rangle$ is then,
\be
\langle \dot{E}(t) \rangle =   - \langle \sum_{j=1}^M \Delta E_j
\delta(t-\sum_{k=0}^{i-1} 
\tau_k)\rangle_{\DE, \tau} \;\;\;.\label{Edot}
\ee
\begin{figure}
\begin{center}
\leavevmode
\psfig{figure=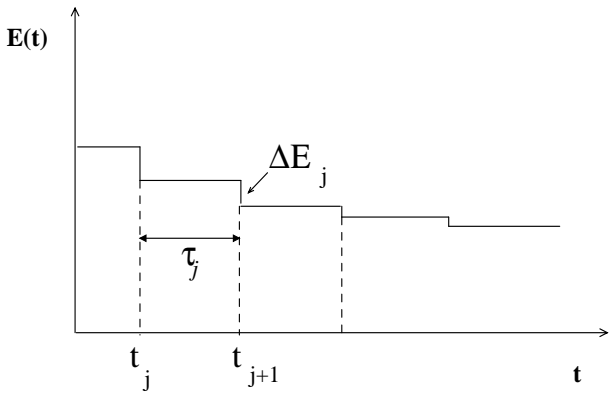,width=8cm,height=5cm,angle=270}
\end{center}  
\narrowtext   
\caption{
 A schematic plot of the variation of the total energy with time.}
\label{fig4}
\end{figure}
The distribution of waiting times $\tau_k$ is 
dependent only on the configuration of the chain at the $k$' th step and 
independent of the previous waiting times.  
Since the dynamics is just changing a pair of dihedral angles in opposite 
directions, for each conformation $\{ \phi_i \}$ one may define an associated 
chain of $N(N-1)/2$ sites, with each site corresponding to a pair $(i, 
i^\prime)$ on the original chain. On the associated chain, a site will be
assigned the value 1 if the corresponding pair has at least one 
``allowed'' move, and the value 0 if both moves are ``blocked.''
Now the probabilities of encountering allowed or blocked moves as one
implements the Metropolis dynamics outlined above are simply given by
the density of 1' s or 0' s on the associated chain at a given
relaxation step, namely,
$p_k$ and $q_k=1-p_k$. Therefore, in the $k$'th conformation,
the probability of making a transition after $\tau_k$ blocked 
moves simply obeys the first passage time distribution~\cite{Feller}, 
\be
P_k(\tau_k)=\mu_k e^{-\mu_k \tau_k} \;\;\;,\;\;\; \mu_k \equiv \mathopen|
\ln q_k \mathclose| \;\;\;.
\ee
\noindent
Writing the $\delta$ function in equation~(\ref{Edot}) in the Fourier representation
and performing the $\tau$-integrals we get
\be
\langle \dot{E} (t) \rangle  = 
\frac{1}{2\pi}\sum_{j=1}^M\left<\DE_j
 \sum_{\ell=1}^{j-1}\prod _{{k=0} \atop{k\neq \ell}}^{j-
1} \left( \frac{\mu_{k}}
{\mu_{k}-\mu_{\ell}}\right)e^{-\mu_{\ell} t} \right>_{\DE}\;\;\;.
\ee
We may argue that the larger the energy loss in a relaxation event, the
longer it will take for the phase point to make a transition out of this state. 
Since $\mu_k$ is roughly the 
expectation for $\tau_k$, we assume that
$\mu_k \sim 1/\DE_k$. 
With the assumption that the energy steps encountered along a relaxation
path are independently distributed, i.e., 
$P(\Delta E_1 \ldots \Delta E_M) = \prod_{s=1}^{M} P(\DE_s)$ for a process
of $M$ steps, one finds,
\be
\langle \dot{E}(t) \rangle =
-\frac{1}{2\pi} \sum_{j=1}^M \langle \DE_j \rangle \sum_{\ell=1}^{j-1} I_{j \ell}
(t) \;\;\;, 
\label{E2}
\ee
\noindent
where $I_{j,\ell}(t)$ is
\bea
I_{j,\ell}(t) \equiv \int_0^{\infty} d(\DE_{\ell} & ) &
e^{- \frac{t}{\DE_{\ell}}} P(\DE_{\ell}) \cr
& \times & \left[ \prod _{{k=0} 
\atop{k\neq \ell}}^{j-1} \left< \frac{\DE_{\ell}}{\DE_{\ell}-\DE_{k}} 
\right>_{\DE_k}
\right] \;\;\;\; . \label{Ij} \eea
\noindent
We have determined from our simulations that the distribution $P(\Delta E_\ell)$  
(see Fig. 5) also has the stretched exponential form  
$P(\DE_\ell)= P_o e^{-(\Delta E_\ell)^\alpha }$.
The angular brackets then take the form 
\be
\Delta E_{\ell} \int_{0}^{\infty}
(\Delta E_{\ell}-\Delta E_k)^{-1} \exp (-({\Delta E_k})^\alpha) d\Delta E_k
\ee
\noindent
which we approximate by $\Delta E_{\ell} \exp(-({\Delta E_k})^\alpha)$. The 
integration in equation (\ref{Ij}) is then straightforward, leading, upon 
substitution in (\ref{E2}), to
\be
E(t) \sim t \sum_{j=1}^{M} \left( \frac{j-1}{j} \right) \exp(-a_j t^\beta)
\ee
\noindent
where $a_j = j (1-\alpha) (\alpha j)^{-\beta} (1+\beta)^{-1}$ and
\be
\beta=\frac{\alpha}{\alpha+1} \;\;\;.
\ee
\noindent
Substituting the value
of $\alpha$ we find from our 
simulations, namely $\alpha=0.39 \pm 0.02$, we get $\beta = 0.28 \pm 0.01$ 
which is the result we obtained from the 
fits to the MC simulations within our error bounds.

\begin{figure}
\begin{center}
\leavevmode
\psfig{figure=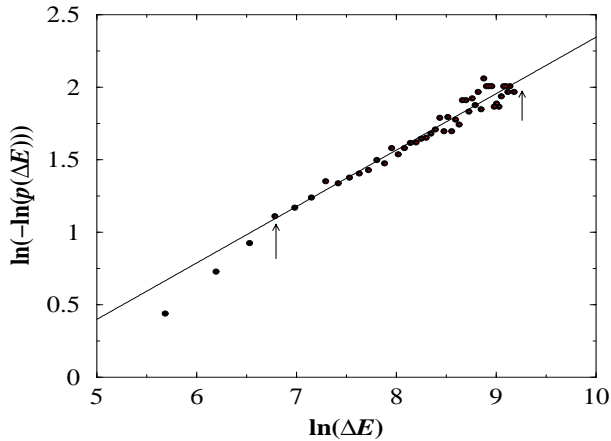,width=8cm,height=6cm,angle=0}
\end{center}
\narrowtext
\caption{The distribution of energy differences encountered along the
relaxation path are fit
to a stretched exponential. Level spacing histograms were formed for
chains of $N=48$ and averaged over 100 runs for the
zero-temperature Metropolis  
relaxation. The exponent $\alpha$ of the stretched exponential is found to
be $0.39 \pm 0.02$.}
\label{step}
\end{figure} 

A study of the correlations between
fluctuations about the native state~\cite{Erman3}
for the Beads-and-Springs model,
with the H-P sequence  and the contact map for the native states for seven
real proteins (6lyz, 1cd8, 1bet, 1fil, 1bab, 1csq and 1hiv) was performed
by Erman~\cite{Erman4}.
Using the Gaussian approximation~\cite{Ewen} to the coherent scattering
function and a normal mode analysis~\cite{Tirion,Bahar,Go,Levitt},
he also  finds a stretched exponential relaxation with $\beta=1/4$.
Experiments on real proteins and
polymers~\cite{Wolynes1,Angell,Bahar1,Colmenero}
yield $0.2\le\beta\le 0.4$.
Our  results seem to be  closer to 1/4 and smaller than the values
most commonly found for spin-glasses~\cite{Mezard}, namely 1/3.
It should also be noted that glassy behaviour is obtained here in the
absence of quenched randomness, or of frustration arising from steric hindrances,
which we do not take into account.

Comparing the relaxation behaviour near the native state
 with the behaviour we observe at relatively high
energies for random heteropolymers,
we conclude that the relaxation behaviour, and therefore 
the dynamics and the structure of the energy landscape  are universal
over a very
large range of energies, and 
are relatively independent of the specific sequence or the details
of the dynamics.

We are grateful to Burak Erman for motivating this work and for the
close interest he has shown in every stage of its progress and to 
Mustansir Barma for a useful discussion. One of us
(A.E.) acknowledges partial support by the Turkish Academy of Sciences.

\end{multicols}

\begin{thebibliography}{20}
\bibitem{Wolynes1}  H. Frauenfelder, S. G. Sligar, P. G. Wolynes, 
Science, {\bf 254}, 1598 (1991).
\bibitem{Wolynes2} P.G. Wolynes, J.N. Onuchic, D.Thirumalai,
Science, {\bf 276}, 1619 (1995)
\bibitem{ben-Avraham} D. ben-Avraham, Phys. Rev. B {\bf 47}, 14
559 (1993).
\bibitem{Dill} K.A. Dill, S. Bromberg, K. Yue, K.M. Feibig, D.P.
Yee, P.D. Thomas and H.S. Chan, Protein Science {\bf 4}, 561
(1995).
\bibitem{Tirion} M.M. Tirion, Phys. Rev. Lett. {\bf 77}, 1905
(1996).
\bibitem{Bahar} T. Haliloglu, I. Bahar, and B. Erman, Phys. Rev.
Lett. 79, 3090 (1997).
\bibitem{Erman1} B. Erman and K. Dill, J.Chem. Phys., in press.
\bibitem{Erman2} B. Erman, ``Hydrophobic Collapse of Proteins into
their Near-Native Configurations," unpublished.
\bibitem{Tuzel} E. T\"uzel, A. Erzan, to be published.
\bibitem{Angell} J.L. Green, J. Fan, and C.A. Angell, J.
Phys.Chem. {\bf 98} 13780 (1994).
\bibitem{Bahar1} B. Erman, I. Bahar, Macromol.Symp. {\bf 133},
33 (1998).
\bibitem{Colmenero} J. Colmenero, A. Arbe, and A. Alegria, 
Phys. Rev. Lett. {\bf 71}, 2603 (1993).
\bibitem{Tang} R. M\'elin, H. Li, N. S. Wingreen and C. Tang, J. Chem. Phys.
{\bf 110} 1252 (1999).
\bibitem{Flory} P.J. Flory, {\it Statistical Mechanics of Chain Molecules},
  (Interscience, N.Y., 1969).
\bibitem{Feller} W. Feller, {\it An Introduction to Probability
Theory and its Applications} (Wiley, N.Y. 1957), Vol. I and II.
\bibitem{Erman3} B. Erman, J. Comp. Polym. Sci., in press.
\bibitem{Erman4} B. Erman, private communication.
\bibitem{Ewen} B. Ewen, D. Richter, in {\it Elastomeric Polymer Networks}, J. E.
Mark and
B. Erman ed., (Prentice Hall, 1992) pp. 220.
\bibitem{Go} N. Go, T. Noguti, T. Nishikawa, Proc. Natl. Acad. Sci. (USA) {\bf 80}, 
3696 (1983).
\bibitem{Levitt} M. Levitt, C. Sander, and P. S. Stern, J. Mol. Biol. {\bf 181}, 423
(1985).
\bibitem{Mezard} M. Mezard, G. Parisi and M. A. Virasoro,
{\it Spin Glass Theory and Beyond} {World Scientific, Singapore 1987}.
\end{thebibliography}
\end{document}